# Random matrix route to image denoising

Gaurab Basu[a], Kaushik Ray[b], Prasanta K. Panigrahi[c*]

[a]*NIT Durgapur, Durgapur713209, West Bengal, India*

[b]*Meghnad Saha Institute of Technology, Uchhepota Kolkata-700150, West Bengal, India*

[c]*Indian Institute of Science Education and Research- Kolkata, Mohanpur Campus, Mohanpur - 741252, West Bengal, India*

**Abstract**

We make use of recent results from random matrix theory to identify a derived threshold, for isolating noise from image features. The procedure assumes the existence of a set of noisy images, where denoising can be carried out on individual rows or columns independently. The fact that these are guaranteed to be correlated makes the correlation matrix an ideal tool for isolating noise. The random matrix result provides lowest and highest eigenvalues for the Gaussian random noise for which case, the eigenvalue distribution function is analytically known. This provides an ideal threshold for removing Gaussian random noise and thereby separating the *universal* noisy features from the *non-universal* components belonging to the specific image under consideration.

## 1. Introduction

The fact that statistical error is reduced on averaging can be effectively used for denoising of images [1]. However, in certain cases, it may be of interest to isolate the noise from the physical features, in order to analyze the properties of the same. For example, in chaos based cryptography, the image is hidden in a chaotic output generated from a deterministic dynamical system, whose character significantly differs from random noise [2]. In the context of denoising, several methods have been proposed to find out a threshold to separate noise from the image. The fact that image features are highly structured and as in many cases, can be captured by suitable mathematical functions, which is not the case for random noise, forms the basis for estimating this threshold. For example, the image decomposition in the wavelet domain can isolate image features, showing scale preference, from noise, which shows identical behavior at different scales. This has led to the well-known Donoho universal threshold [3], which is implemented at every level of the wavelet decomposition. Donoho's soft thresholding method has been improved over the years. The descriptive approaches have been replaced by discriminative approach, yielding better results [4]. Recent wavelet based methods include curvelets [5], complex wavelets [6], steerable pyramids [7] and scale mixtures in the wavelet domain [8].

Singular value decomposition (SVD) of images has also been used for this purpose. On physical grounds, it can be expected that in the correlation matrix involving images, image features will be stored in the dominant eigenvalues and the noise in a separate domain of lower eigenvalues. This has formed the basis for identifying the thresholds for separation of noise from image features. Based on the above principle, new algorithms include redundant representation of image patches and then denoising them in the least square sense[9], as also block based noise estimation techniques[10]. Recently wavelets and SVD based approaches have been combined, where SVD has been applied to various high pass sub bands of wavelet decomposition, for extracting a threshold at each level [11]. However, the lack of a precise mathematical difference between noise and correlated features has made determination of threshold value subjective [12].

In the present paper, we make use of recent results in random matrix theory to isolate Gaussian white noise, the so called universal part, from the non-universal image

*Corresponding author, Mob.: +91 9748918201.

E-mail addresses: pprasanta@iiserkol.ac.in, (P.K. Panigrahi).

features [13]. This is carried out through the study of eigenvalues of the correlation matrix. The above result from the random matrix theory [14], identifies the minimum and maximum eigenvalues of a correlation matrix involving Gaussian random noise, as also the probability distribution function (PDF) of these eigenvalues. This significant result has found application in isolating the universal noise from non-universal characteristic features of financial time series [15], atmospheric data [15], spectral fluctuations of cancer and normal tissue fluorescence data [16, 17]. Here, we make use of this procedure to mathematically identify a precise threshold, which can separate Gaussian random noise from features in an image. For the present algorithm to work, it requires an ensemble of similar pictures with noise. This requirement is readily fulfilled for cases like satellite and medical images, where multiple copies of a given image exist. In that case, one can analyse copies of a given row or column of an image to construct a correlation matrix, and make an effective isolation of noise from the image features.

## 2. Approach

As has been mentioned earlier, for Gaussian random noise, the corresponding correlation matrix has lowest and highest eigenvalues, between which the other eigenvalues are distributed obeying an analytically known distribution function. The maximum and minimum eigenvalues, as well as the distribution function depends on the size of the correlation matrix and the width of the Gaussian random noise.

The correlation matrix is given by:

$$C_{ij} = \frac{1}{K}\sum_{k=1}^{K} \delta x_i(k)\delta x_j(k).$$

Here K=length of the matrix i.e., the number of columns of the given image; $\delta x_i(k)$ refers to the mean subtracted pixel value of the i$^{th}$ row and k$^{th}$ column of a given image, from the collection of noisy images [18].The previous equation can also be written in the form:

$$C = \frac{1}{K}MM^T;$$

here M is a $N \times K$ rectangular matrix, which is formed by collecting a particular row of all the images from the available noisy image set. It has been found that the density of the eigenvalues of the correlation matrix $\rho c(\lambda)$, defined as,

$$\rho c(\lambda) = \frac{1}{N}\frac{dn(\lambda)}{d\lambda}$$

and $\rho c(\lambda)$ is also given by:

$$\rho c(\lambda) = \frac{Q}{2\pi\sigma^2}\frac{\sqrt{(\lambda_{max}-\lambda)(\lambda-\lambda_{min})}}{\lambda},$$

where

$$\lambda_{min}^{max} = \sigma^2\left(1 + \frac{1}{Q} \pm 2\sqrt{\frac{1}{Q}}\right).$$

$n(\lambda)$ is the number of eigenvalues of C less than λ and N refers to the number of snapshots i.e., number of images in the noisy set. Here λ belongs in the interval $[\lambda_{max}, \lambda_{min}]$ and Q=K/N. It is worth mentioning that in case of an N×K random matrix M, the above exact distribution of the eigenvalues has been derived in the limits K→∞ and N→∞, with Q≥1. For N being finite there is a small probability of finding eigenvalues above or below the specified and calculated range of $[\lambda_{max}, \lambda_{min}]$, which generally tends to zero as the size of the matrix M increases.

As is evident from above, $\lambda_{max}$ gives an ideal handle for designing a threshold to remove the eigenvalues, corresponding to noise and reconstruct the physical component of the data from the remaining eigenvalues [18, 13].

For the purpose of illustrating the efficacy of the proposed algorithm, we start with the Lena image of size 200× 200, as given in Fig 1(a). Hundred copies of this image are produced; each placed with Gaussian random noise of strengths 50, 100 and 200.The noise strength is estimated from the standard deviation of the noise matrix being added to the image.

A sample from each set is shown in Fig. 1(b), 1(c) and 1(d) respectively. Each row is then denoised independently. A matrix M of size 100x200 is constructed, from the rows of the noisy set as described earlier, which was subsequently normalized by mean subtraction and division by width.

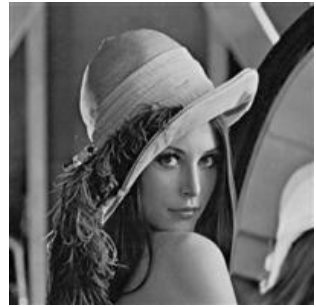 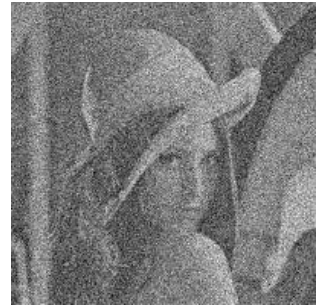

Fig. 1(a).    Fig. 1(b).

---

* The noise strength is estimated from the standard deviation of the noise matrix being added to the image.

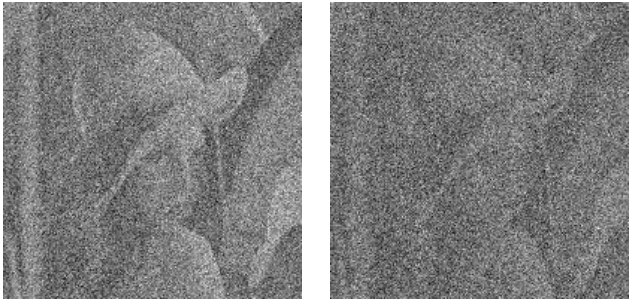

Fig. 1(c).                Fig. 1(d).

Fig. 1(a) shows the original Lena image and the following images show the same with added noise of different strengths of 50*, 100* and 200* respectively.

It should be noted that when the Gaussian random matrix is square matrix, the corresponding correlation matrix eigenvalues obey the Wigner's semicircle [13] law for which the density has a singularity at origin. In case of non-square matrix the corresponding density is non-singular. To show the appropriateness of our procedure, we have considered two cases of given image one corresponding to a square matrix (image size 200 with 200 snapshots) and other a non-square matrix (image size 200 with 100 snapshots). Fig. 2 depicts the histogram of eigenvalues remarkably well fitted with corresponding exact probability densities.

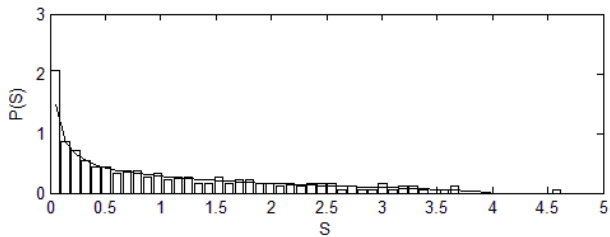

Fig. 2(a).

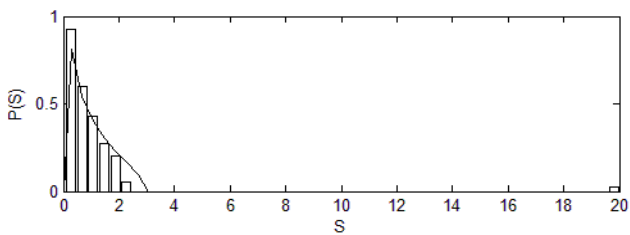

Fig.2(b).

Fig. 2(a) and (b) show the histogram of the eigenvalues of the correlation matrix of images with snapshots 200, 100 with noise 500 and 200 respectively, fitted with the corresponding probability density function. As expected the increase in the strength of the noise takes the eigenvalue distribution between $\lambda_{max}$ and $\lambda_{min}$ closer to the analytical forms.

The noise can be eliminated by removing all the eigenvalues lying between $\lambda_{max}$ and $\lambda_{min}$. The histogram of the eigenvalues is shown in Fig 2(a) and 2(b). The unfolding operation removes the trend after which the eigenvalues can be compared with random matrix prediction. It is observes that large number of eigenvalues lie between ($\lambda_{max}$ and $\lambda_{min}$), with an eigenvalue distribution matching reasonably well with the derived one. We have checked with a number of simulations that for finite sized matrices derived from Gaussian random entries all the eigenvalues lie between ($\lambda_{max}$ and $\lambda_{min}$) although the distribution may differ slightly from the derived one. Distribution fits better as the matrix size increases. $\lambda_{max}$ is treated as a threshold and all the eigenvalues below it are removed. The remaining eigenvalues are then used to reconstruct the rows of the image. It is worth pointing out that in many images only one eigenvalue survived after thresholding. This is physically expected since the given rows or columns of one image are expected to be highly correlated. This operation is then repeated for all the columns of the image with different strengths of the noise, for which denoising works reasonably well. It is clearly observed that, in many cases the image is invisible due to the noise. The images reconstructed from the denoised columns are shown in the Figs. 3(a), 3(b).

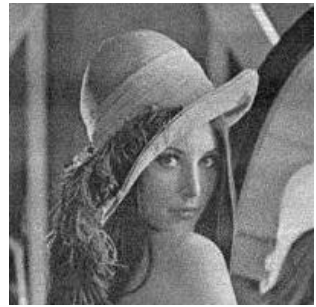 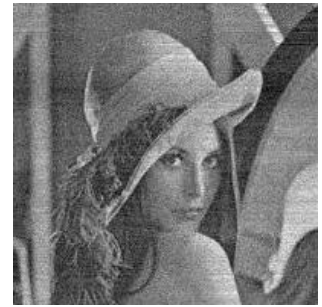

Fig. 3(a).                Fig. 3(b).

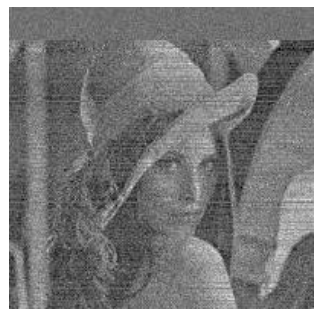

Fig. 3(c).

Fig. 3(a) depicts the reconstructed image from the noisy image setas shown in Fig. 1(b) which was corrupted with the noise strength of 50* and 100 snapshots was used for reconstruction. 3(b) shows the reconstructed image from 1(c) which was corrupted with noise strength of 100* and 100 snapshots was used for reconstruction and 3(c) shows the reconstructed image from 1(d) which was corrupted with noise strength of 200* and 100 snapshots was used for reconstruction.

From the results it is obvious that the noise removal is quite effective. The PSNR,

$$PSNR = 20 \log_{10}\left(\frac{255}{RMSE}\right),$$

$$RMSE = \sqrt{\frac{1}{MN}\sum_{i=0}^{M}\sum_{j=0}^{N}[I(i,j) - \hat{I}(i,j)]^2},$$

is shown in Fig: 4. Here $I$ is the original is image and $\hat{I}$ is the denoised image [8]. It is observed that the PSNR initially drops and then gets saturated after the $80^{th}$ snap shot at around a PSNR value of 23. Similar features were observed for other rows as well. As expected, statistical features become prominent only in an ensemble of reasonably large number of images [19].

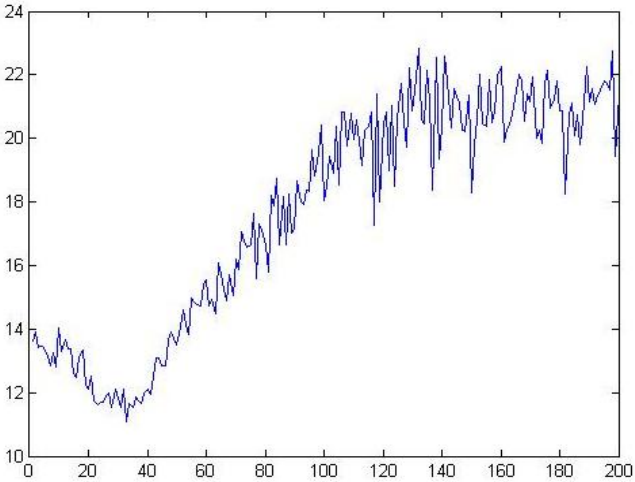

Fig. 4. Depicts the PSNR values along the y axis, number of snapshots along the x axis, with noise strength as 100.

Fig. 5 depicts the PSNR values for different noise strengths, with respect to the number of snapshots. One observes that, if the image is noisier, the PSNR saturation is arrived with less number of snap shots. For these noisy images, it is further found that in certain cases, when the number of snapshots are less, all the eigen values lie between λmax and λmin, this leading to a failure in image reconstruction. This once more illustrates the need to take large number of snapshots for this statistical procedure to work.

## 3. Algorithm

In this section we describe the proposed algorithm in a step by step manner.

1. First N samples of an image are assembled which are corrupted by Gaussian random noise,
2. As per the algorithm, denoising of each row has to be done separately.
3. In order to denoise a particular $i^{th}$ row
3.1. Form a matrix comprising of the $i^{th}$ rows of the various image samples.
3.2. Singular value decomposition of the normalized matrix is then carried out.
3.3. Subsequently a threshold is applied on the eigenvalues matrix based on the random matrix result.
3.4. We reconstruct the matrix of the rows using the eigenvalues greater than λmax and extract the denoised row.
4. We reconstruct the image from each of these denoised rows.

## 4. Results

We have applied the above algorithm on many standard test images, e.g., Lena, Baboon, House and Peppers etc. We have taken different number of snapshots and images of different noise strengths, for the same image and tabulated the results in Table 1. It is observed that for lower noise strengths, PSNR saturation is achieved with a smaller ensemble of images and an optimal value of PSNR is obtained for medium noise strengths. As noise strength is further increased, PSNR does not improve substantially even if the number of snapshots are increased.

Table 1

Summary of results for different images.

| Image name | Noise Strength (Gaussian)* | PSNR | Time Taken(s)** | Image Size | Number of snapshots |
|---|---|---|---|---|---|
| Lena | 49.60 | 15.9892 | 2.77 | 200x200 | 50 |
| | | 25.2708 | 23.47 | 200x200 | 200 |
| | 100.33 | 21.7004 | 7.65 | 200x200 | 100 |
| | | 22.4001 | 24.25 | 200x200 | 200 |
| | 200.45 | 18.4354 | 14.91 | 200x200 | 150 |
| | | 19.0801 | 22.96 | 200x200 | 200 |
| Baboon | 49.60 | 18.3768 | 2.83 | 200x200 | 50 |
| | | 20.6930 | 23.07 | 200x200 | 200 |
| | 100.33 | 21.5330 | 7.62 | 200x200 | 100 |
| | | 18.0715 | 23.11 | 200x200 | 200 |
| | 200.45 | 19.2988 | 15.19 | 200x200 | 150 |
| | | 18.8594 | 23.82 | 200x200 | 200 |
| House | 49.60 | 15.6165 | 1.80 | 180x180 | 50 |
| | | 18.4693 | 15.61 | 180x180 | 200 |
| | 100.33 | 17.8550 | 4.32 | 180x180 | 100 |
| | | 17.9017 | 15.97 | 180x180 | 200 |
| | 200.45 | 15.1243 | 8.72 | 180x180 | 150 |
| | | 17.2433 | 15.61 | 180x180 | 200 |
| Peppers | 49.60 | 16.0311 | 1.73 | 180x180 | 50 |
| | | 19.3171 | 15.64 | 180x180 | 200 |
| | 100.33 | 16.6684 | 4.60 | 180x180 | 100 |
| | | 17.3761 | 15.94 | 180x180 | 200 |
| | 200.45 | 17.1662 | 8.97 | 180x180 | 150 |
| | | 17.2389 | 15.70 | 180x180 | 200 |

*The noise strength is estimated from the standard deviation of the noise matrix being added to the image.

**The processing has been carried out in a standard AMD 5200+ processor based computer with 1GB of ram.

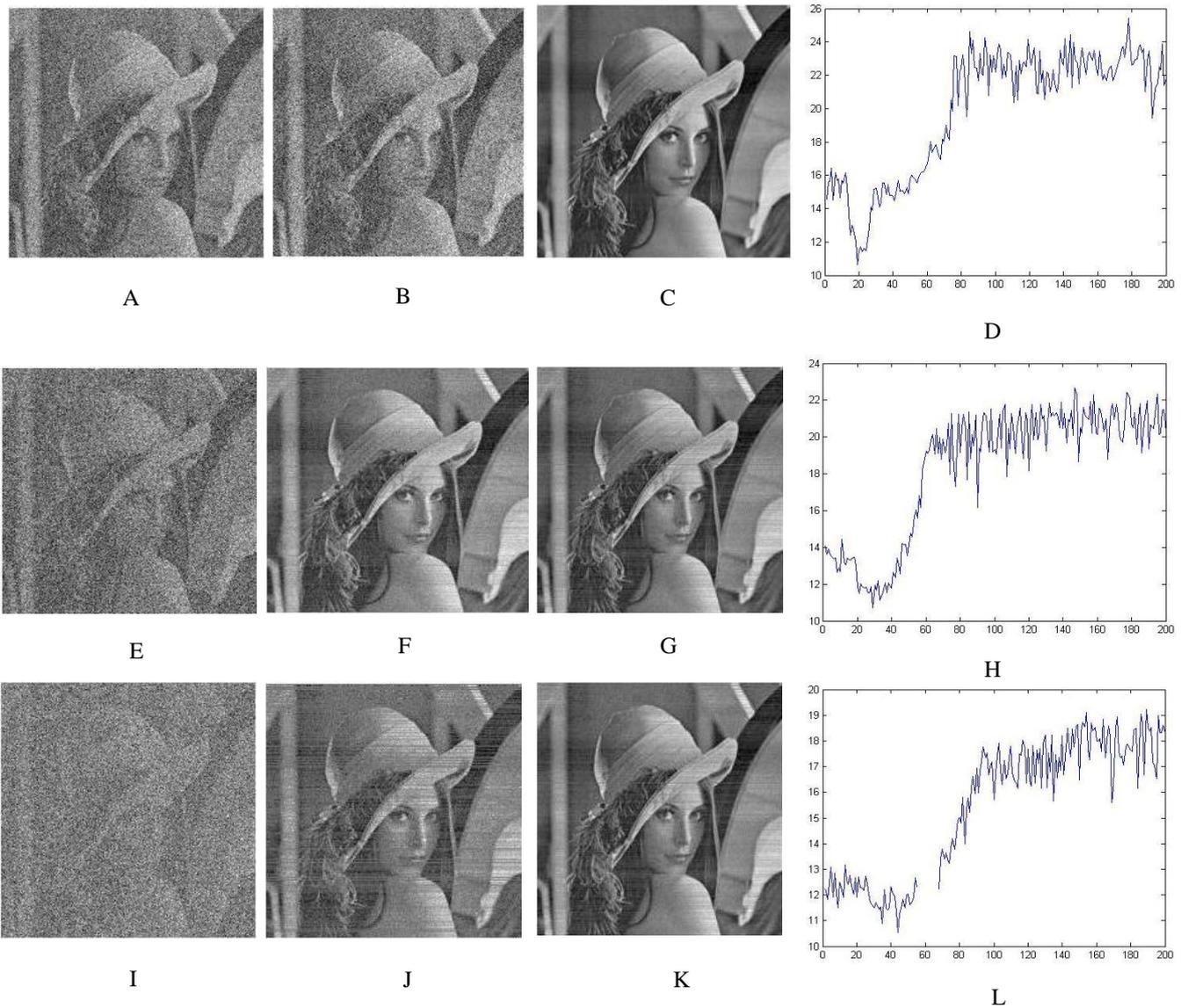

Fig. 5. Depicts: (A) Lena with noise strength 50, and (B) denoised Lena using 50 snapshots, (C) denoised Lena using 200 snapshots. (D),(H) and (L) show the corresponding PSNR plots, with 'x' axis showing the number of snapshots and 'y' axis corresponding PSNR values. (E) depicts Lena corrupted with Gaussian random noise of strength 100; (F) shows the corresponding denoised Lena, with 100 snapshots, (G) shows the same denoised with 200 snapshots. (I) depicts Lena corrupted with Gaussian random noise of strength 200, (J) depicts the corresponding denoised Lena with 150 snapshots; (K) shows the same with 200 snapshots.

## 5. Conclusion

In conclusion, we have illustrated a denoising procedure based on the eigenvalue structure of the correlation matrices. Our method makes use of the fact that for correlation matrices derived from Gaussian random numbers, it is possible to exactly find the lowest and highest eigenvalues, with an analytically defined probability function, characterizing the eigenvalue distribution. The fact that eigenvalue structure of Gaussian noise is well defined in the correlation domain makes the same ideal for separating noise from the structured image features. The physically tenable assumption that the noise and image features, having significantly different correlation properties will occupy well separated domains in the eigenvalue spectra makes $\lambda_{max}$ a suitable candidate for thresholding. For the proposed algorithm to work for denoising, one needs to have multiple copies of the given noisy image. The fact that in a number of cases multiple images of the same object may be available makes this algorithm well suited for the same. We have demonstrated through a number of examples that the image can be extracted with good accuracy. Even for extremely noisy images, it was found that the present procedure extracts the image features with reasonable accuracy.